\documentclass[
 reprint,
 showpacs, preprintnumbers,
 amsmath, amssymb,
 aps,
 pra,
 floatfix,
 superscriptaddress
]{revtex4-2}

\usepackage{graphicx}
\usepackage{tikz}
\usepackage{xcolor}
\usetikzlibrary{arrows}

\makeatletter
\newcommand{\colorcaption}[2][]{%
  \begingroup%
  \renewcommand{\@caption@fignum@sep}{ (Color online). }%
  \caption[#1]{#2}%
  \endgroup%
}
\makeatother

\makeatletter
\def\maketag@@@#1{\hbox{\m@th\normalfont\normalsize#1}}
\makeatother

\usepackage{dcolumn}
\usepackage{bm}
\usepackage{bbm}
\usepackage{upgreek}
\usepackage{hyperref}
\usepackage{blindtext}
\usepackage{hyphsubst}

\usepackage{changes}

\usepackage{dsfont}

\usepackage{newtxtext,newtxmath}

\definecolor{darkgreen}{RGB}{0 100 0}

\begin{document}

	\preprint{}


	\title{Transport Signatures of Radial Rashba Spin-Orbit Coupling \\ at Ferromagnet/Superconductor Interfaces}
	
	\author{Andreas Costa}%
	\email[Corresponding author: ]{andreas.costa@physik.uni-regensburg.de}
 	\affiliation{Institute for Theoretical Physics, University of Regensburg, 93040 Regensburg, Germany}

 	\author{Jaroslav Fabian}%
 	\affiliation{Institute for Theoretical Physics, University of Regensburg, 93040 Regensburg, Germany}

 	\date{\today}

    
    \begin{abstract}
        Spin-orbit coupling~(SOC) emerging at the interfaces of superconducting magnetic tunnel junctions is at the heart of multiple unprecedented physical phenomena, covering triplet proximity effects induced by unconventional (spin-flip) Andreev reflections, giant transport magnetoanisotropies, sizable tunneling anomalous Hall effects, and electrically controlled current-reversing $ 0 $--$ \pi $(-like) transitions in Josephson contacts. 
        Recent first-principles calculations proposed that the Rashba spin-orbit fields in twisted graphene/transition-metal dichalcogenide and van der Waals multilayers can---owing to broken mirror symmetries---exhibit an unconventional radial component~(with spin parallel to the electron's momentum), which can be quantified by the Rashba angle~$ \theta_\mathrm{R} $. 
        We theoretically explore the ramifications of radial Rashba SOC at the interfaces of vertical ferromagnet/superconductor tunnel junctions with a focus on the magnetoanisotropies of the tunneling and tunneling-anomalous-Hall-effect conductances. 
        Our results demonstrate that $ \theta_\mathrm{R} $ can be experimentally extracted from respective magnetization-angle shifts, providing a robust way to probe the radial Rashba SOC induced by twisted multilayers that are placed as tunneling barriers between ferromagnetic and superconducting electrodes. 
    \end{abstract}
    
    
    \maketitle
    

    \section{Introduction  \label{sec:1}}

    Spin-orbit coupling~(SOC) is essential for electrically manipulating electron spins in spintronics devices~\cite{Fabian2004,Fabian2007}. 
    Tunnel junctions that consist of different-material electrodes are particularly interesting to explore~\cite{MatosAbiague2009,MatosAbiague2015,Hoegl2015,Hoegl2015,*Hoegl2015a,Costa2017,Vezin2020,Mondal2024}, as their inherently broken space-inversion symmetry induces strong interfacial Rashba SOC~\cite{Bychkov1984,Bychkov1984b,*Bychkov1984c}. 
    If the tunneling barriers furthermore lack bulk-inversion symmetry---like, e.g., zinc-blende semiconductors~\cite{Gmitra2013}---the Rashba spin-orbit fields interfere with additionally emerging spin-orbit fields of the Dresselhaus type~\cite{Dresselhaus1955}, and unique transport features such as the $ C_{2v} $-symmetric tunneling anisotropic magnetoresistance effect~\cite{Moser2007,MatosAbiague2009}---referring to in-plane tunneling-conductance magnetoanisotropies---or the tunneling anomalous Hall effect~\cite{MatosAbiague2015,Rylkov2017} appear.

    Two-dimensional materials and van der Waals multilayer structures are attracting enormous research interest~\cite{Han2014,Zutic2019,Avsar2020,Sierra2021,Perkins2024}, as many of their physical properties can be tuned by gating, twisting, the number of monolayers, or their stacking order~\cite{Gmitra2016,Li2019,David2019,Naimer2021,*Naimer2021a,Peterfalvi2022,Veneri2022,Lee2022,Zollner2023a,Naimer2023,Zollner2023,Naimer2024}. 
    Proximitizing graphene by transition-metal dichalcogenides was, for example, predicted to induce novel Ising-like valley--Zeeman SOC that results in giant spin-relaxation anisotropies~\cite{Cummings2017,Ghiasi2017,Zihlmann2018,Benitez2018}. 
    The broken space inversion in graphene/transition-metal-dichalcogenide and van der Waals multilayers likewise generates Rashba spin-orbit fields. 
    Pioneering first-principles studies proposed that breaking mirror symmetries by twisting the monolayers can---apart from the conventional Rashba with a spin texture that is perpendicular to the electron momentum---imprint a (purely) unconventional radial Rashba component, which prefers a spin alignment parallel to the momentum, on the Rashba spin-orbit field~\cite{Li2019,David2019,Menichetti2023,Frank2024}; the ratio between conventional and radial Rashba SOCs is quantified by the Rashba angle $ \theta_\mathrm{R} $~($ \theta_\mathrm{R} = 0 $ indicates conventional Rashba SOC, $ \theta_\mathrm{R} = 0.5\pi $ radial Rashba SOC, and $ 0 < \theta_\mathrm{R} < 0.5 \pi $ an admixture of both).

    Extremely rich physics is bound to occur when SOC interacts with ferromagnetism and superconducting coherence~\cite{Eschrig2011,Linder2015}. 
    Among the perhaps most striking phenomena are unconventional spin-flip Andreev reflections that induce superconducting triplet correlations in proximitized regions~\cite{Bergeret2001,Volkov2003,Keizer2006,Halterman2007,Eschrig2008,Eschrig2011,Sun2015,Costa2021} and lead to marked transport anomalies, such as remarkably enhanced conductance magnetoanisotropies~\cite{Hoegl2015,*Hoegl2015a,Jacobsen2016,Costa2017,Martinez2020,Vezin2020}---termed magnetoanisotropic Andreev reflection~(MAAR)---or sizable superconducting tunneling anomalous Hall effects~\cite{Costa2019}, as well as the recently intensively investigated supercurrent diode effect~\cite{Ando2020,Baumgartner2022a,Baumgartner2022,Costa2023,Costa2023a,Banerjee2023,Kochan2023,Banerjee2024,Banerjee2024a,Kokkeler2024,Reinhardt2024,Kang2024,Scharf2024}.

    In Ref.~\cite{Kang2024}, the authors performed numerical transport calculations for lateral graphene-based van der Waals heterostructures and identified robust magnetotransport signatures to disentangle conventional and radial Rashba  SOCs in experiments. 
    Furthermore, Ref. \cite{Costa2024} reported that the interplay of crossed conventional and radial Rashba SOCs at the interfaces of ballistic superconductor/ferromagnet/superconductor Josephson junctions can induce the so-called unconventional supercurrent diode effect. 
    Contrary to the well-studied conventional supercurrent diode effect, which typically originates from a finite center-of-mass momentum that an in-plane magnetic field aligned \emph{perpendicular to the current} imprints on the Cooper pairs~\cite{Daido2022,Yuan2022,He2022,Ilic2022,Davydova2022,Banerjee2023}, the unconventional supercurrent diode effect only occurs when the magnetization of the ferromagnet has a nonzero component perpendicular to the plane of the SOCs, i.e., \emph{parallel to the current}. 
    We unraveled that the spins of the current-carrying electrons are initially polarized in the plane by the spin-orbit field at the left junction interface, travel afterwards through the ferromagnet and precess in this plane around the axis defined by the perpendicular magnetization, and then arrive at the right interface at which they see the second spin-orbit field rotated by an effective angle determining the transmission probability into the second superconductor. 
    If both SOCs are functionally different~(i.e., they are described by different Rashba angles), the rotation angles---and therefore the transmission probabilities---are different for left- and right-propagating electrons, resulting in polarity-dependent critical currents and hence in the unconventional supercurrent diode effect.

    In this paper, we  address the yet unexplored ramifications of radial Rashba SOC induced at the interface of ballistic  vertical ferromagnet/superconductor junctions on the corresponding tunneling-transport~(and thereby microscopically on the Andreev tunneling) and tunneling-anomalous-Hall-effect characteristics with a focus on identifying experimental signatures to determine the Rashba angle $ \theta_\mathrm{R} $. 
    We will demonstrate that the presence of unconventional radial Rashba spin-orbit fields can be probed through the in-plane magnetoanisotropies of the tunneling-anomalous-Hall-effect conductances and Hall supercurrent responses that will both reflect magnetization-angle shifts proportional to $ \theta_\mathrm{R} $~(in agreement with the general symmetry analysis provided in~Ref.~\cite{Kang2024}), or similarly also from the in-plane tunneling-conductance magnetoanisotropies~(i.e., from the in-plane magnetoanisotropic Andreev reflection) when radial Rashba SOC interferes with a functionally distinct spin-orbit field like, for instance, Dresselhaus SOC.

    The paper is organized in the following way. 
    In~Sec.~\ref{sec:2}, we introduce the theoretical model that we apply to compute the tunneling and tunneling-anomalous-Hall-effect conductances and that is kept most general to allow for the simultaneous presence of conventional Rashba, radial Rashba, and Dresselhaus SOCs at the ferromagnet/superconductor junction interface, as well as for arbitrary magnetization directions and bias voltages. 
    We discuss and analyze our numerical results at zero bias in~Sec.~\ref{sec:3}, unraveling the $ \theta_\mathrm{R} $-dependent magnetization-angle shifts as clear transport fingerprints of radial Rashba SOC. 
    In Sec.~\ref{sec:4}, we generalize our model to more realistic junctions, accounting for different chemical potentials or effective masses in the ferromagnetic and superconducting electrodes, and demonstrate that the predicted $ \theta_\mathrm{R} $ shifts are robust features not affected by chemical-potential or mass mismatches. 
    Our findings are summarized in~Sec.~\ref{sec:5}, while Appendix~\ref{sec:AppA} contains additional technical details to calculate the Hall supercurrent responses on the superconducting side of the junction.

    \section{Theoretical Model  \label{sec:2}}

\begin{figure}
        \centering
        \includegraphics[width=0.475\textwidth]{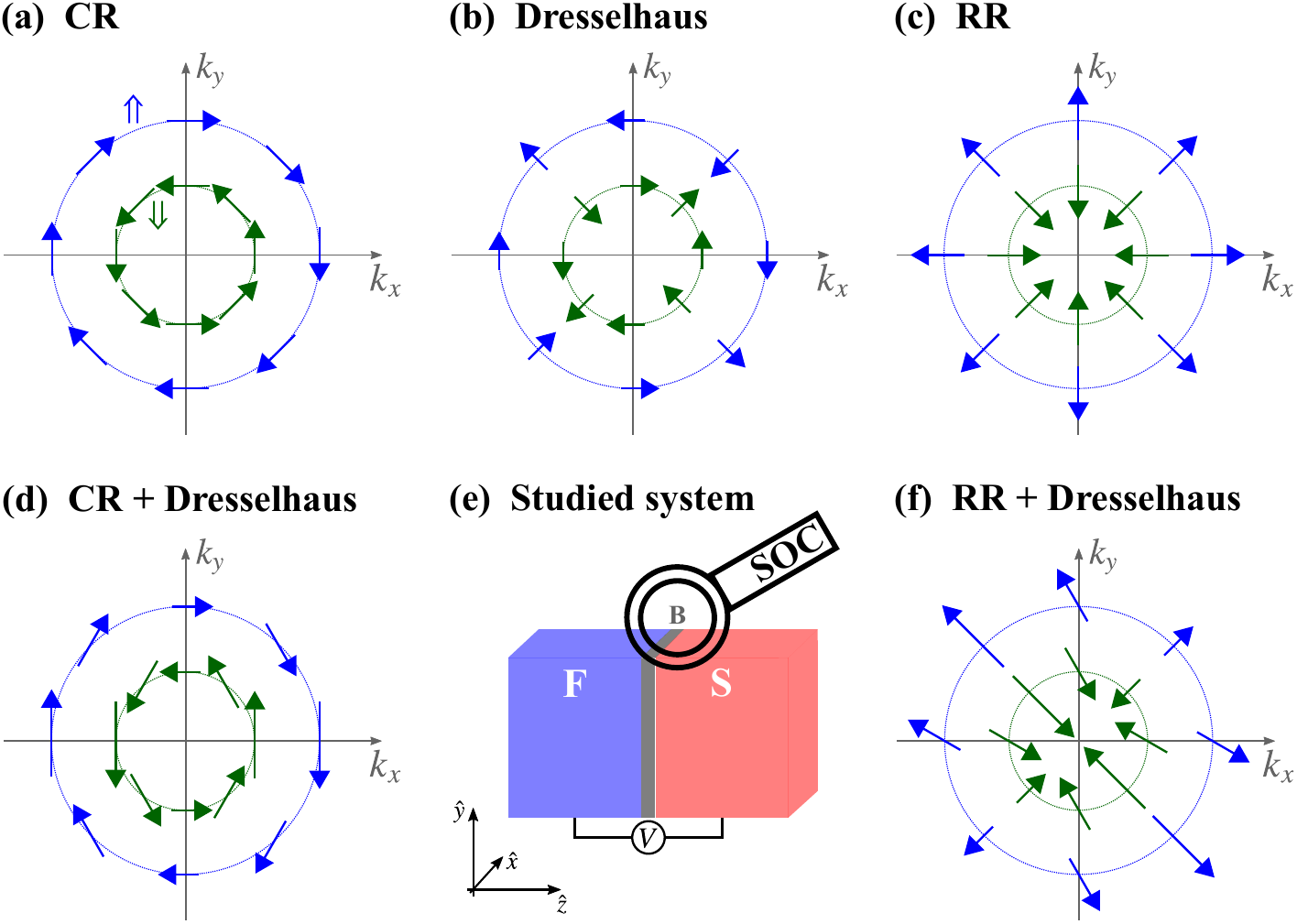}
        \caption{Illustration of the $ \mathbf{k} $-space representation of the spin-orbit fields at the tunneling-barrier interface along the SOC-split spin-up~($ \Uparrow $; blue) and spin-down~($ \Downarrow $; dark-green) Fermi surfaces, defining the preferred in-plane spin orientation, for (a)~conventional Rashba~(CR), (b)~(weak) Dresselhaus, and (c)~radial Rashba~(RR) SOCs. Interference of Dresselhaus with conventional or radial Rashba SOCs results in the spin-orbit fields depicted in~(d) and (f). (e)~Schematical sketch of the considered three-dimensional ferromagnet/barrier/superconductor~(F/B/S) junction using $ C_{2v} $ principal crystallographic orientations $ \hat{x} \parallel [110] $, $ \hat{y} \parallel [\overline{1}10] $, and $ \hat{z} \parallel [001] $. }
        \label{fig:1}
    \end{figure}

    We consider an epitaxial, ballistic, three-dimensional F/B/S tunnel junction, in which the two semi-infinite ferromagnetic~(F; spanning $ z < 0 $) and superconducting~(S; spanning $ z > 0 $) regions are connected by an ultrathin tunneling barrier~(B) at $ z = 0 $; see~Fig.~\ref{fig:1}(e). 
    Additionally to potential scattering, the barrier breaks space-inversion symmetry and induces interfacial SOC.
    Generalizing earlier works~\cite{Hoegl2015,*Hoegl2015a,Costa2017,Costa2019,Vezin2020}, we assume that the corresponding Rashba spin-orbit field can acquire an additional radial component, the ratio of which with respect to the conventional Rashba SOC is determined by the Rashba angle~$ \theta_\mathrm{R} $. 
    A possible experimental realization of the radial Rashba SOC could, for example, exploit a twisted van der Waals bilayer as tunneling barrier.

    To model the junction, we start from its stationary Bogoliubov--de Gennes Hamiltonian~\cite{DeGennes1989}
    \begin{equation}
        \hat{\mathcal{H}}_\mathrm{BdG} = \left[ \begin{matrix} \hat{\mathcal{H}}_\mathrm{e} & \hat{\Delta}_\mathrm{S}(z) \\ \hat{\Delta}_\mathrm{S}^\dagger (z) & \hat{\mathcal{H}}_\mathrm{h} \end{matrix} \right] ,
        \label{eq:BdG}
    \end{equation}
    where 
    $
        \hat{\mathcal{H}}_\mathrm{e} = [ -\hbar^2 / (2m) \boldsymbol{\nabla}^2 - \mu ] \hat{\sigma}_0 - (\Delta_\mathrm{XC} / 2) \Theta(-z) ( \hat{\mathbf{m}} \cdot \hat{\boldsymbol{\sigma}} ) + \hat{\mathcal{H}}_\mathrm{B} 
    $
    refers to the single-electron Hamiltonian and $ \hat{\mathcal{H}}_\mathrm{h} = -\hat{\sigma}_y {\hat{\mathcal{H}}_\mathrm{e}}^* \hat{\sigma}_y $ to its hole counterpart; $ \hat{\sigma}_0 $ ($ \hat{\sigma}_i $) is the $ 2 \times 2 $ identity ($ i $th Pauli spin) matrix and $ \hat{\boldsymbol{\sigma}} = [ \hat{\sigma}_x , \hat{\sigma}_y , \hat{\sigma}_z ]^\top $, while $ \Theta(\ldots) $ indicates the Heaviside step function. 
    The magnetization direction in the ferromagnet is given by the unit vector~$ \hat{\mathbf{m}} = [ \sin(\theta) \cos(\phi) , \sin(\theta) \sin(\phi) , \cos(\theta) ]^\top $---where the polar angle $ \theta $ is measured with respect to the out-of-plane $ \hat{z} \parallel [001] $ axis and the azimuthal angle $ \phi $ with respect to the in-plane $ \hat{x} \parallel [110] $ reference axis---and its exchange splitting is denoted by~$ \Delta_\mathrm{XC} $. 
    The ultrathin barrier is described in a deltalike~[$ \propto \delta(z) $] manner as 
    $
        \hat{\mathcal{H}}_\mathrm{B} = [ V_\mathrm{B} d_\mathrm{B} \hat{\sigma}_0 + \hat{\boldsymbol{\Omega}} (k_x, k_y) \cdot \hat{\boldsymbol{\sigma}} 
        ] \delta(z) , 
    $
    accounting for the barrier height~(width) $ V_\mathrm{B} $~($ d_\mathrm{B} $) and the in-plane spin-orbit field
    \begin{multline} 
        \hat{\boldsymbol{\Omega}} (k_x, k_y) = \Big[  \alpha [ \sin(\theta_\mathrm{R}) k_x + \cos(\theta_\mathrm{R}) k_y ] - \beta k_y , \\
        \alpha [  -\cos(\theta_\mathrm{R}) k_x + \sin(\theta_\mathrm{R}) k_y  ] - \beta k_x , 0 \Big] 
    \end{multline} 
    with general Rashba SOC, quantified by its strength~$ \alpha $ and Rashba angle~$ \theta_\mathrm{R} $, and linearized Dresselhaus SOC parametrized by $ \beta $. 
    Note that $ \theta_\mathrm{R} = 0 $ corresponds to conventional Rashba SOC~[see Fig.~\ref{fig:1}(a)], while $ \theta_\mathrm{R} = 0.5\pi $ represents the unconventional radial Rashba SOC~[see Fig.~\ref{fig:1}(c)]. 
    We additionally include weak Dresselhaus~SOC~(typically assuming $ \beta \ll \alpha $) into our model~[see Fig.~\ref{fig:1}(b)], which could---even in barriers that intrinsically preserve bulk-inversion symmetry---be induced by straining~\cite{Bonell2010} and will be crucial to discern radial Rashba from conventional Rashba SOC in tunneling-transport data~(see~Sec.~\ref{sec:3}). 
    As discussed in Ref.~\cite{Beenakker1997}, we approximate the superconducting pairing potential by $ \hat{\Delta}_\mathrm{S}(z) = \Delta_0 \Theta(z) \hat{\sigma}_0 $ with the zero-temperature superconducting gap~$ \Delta_0 $.
    To simplify the analytical treatment here, we furthermore consider equal chemical potentials $ \mu $ and effective masses $ m $ throughout the junction. The impact of chemical-potential and mass mismatches on our results is briefly analyzed in~Sec.~\ref{sec:4}.

    Since the in-plane wave vector $ \mathbf{k_\parallel} = [k_x, k_y, 0]^\top $ is conserved, the solutions of the Bogoliubov--de Gennes equation $ \hat{\mathcal{H}}_\mathrm{BdG} \Psi^{\sigma} (\mathbf{r}) = E \Psi^{\sigma} (\mathbf{r}) $, where $ \mathbf{r} = [\mathbf{r_\parallel} , z]^\top = [x, y, z]^\top $, factorize into $ \Psi^{\sigma} (\mathbf{r}) = \psi^{\sigma}(z) \, \mathrm{e}^{\mathrm{i} (\mathbf{k_\parallel} \cdot \mathbf{r_\parallel}) } $. 
    The wave functions~$ \psi^{\sigma} (z) $ for the effectively remaining, single-channel, scattering problem perpendicular to the junction interface for incoming electrons with spin~$ \sigma = (-) 1 $, indicating a spin (antiparallel)~parallel to $ \hat{\mathbf{m}} $, read as 
    {\small
    \begin{align}
        \psi^{\sigma} (z<0) &= \mathrm{e}^{\mathrm{i} k_{\mathrm{e}}^{\sigma} z} \chi_\mathrm{e}^{\sigma} 
        + r_\mathrm{e}^{\sigma,\sigma} \mathrm{e}^{-\mathrm{i} k_{\mathrm{e}}^{\sigma} z} \chi_\mathrm{e}^{\sigma}
        + r_\mathrm{e}^{\sigma,-\sigma} \mathrm{e}^{-\mathrm{i} k_{\mathrm{e}}^{-\sigma} z} \chi_\mathrm{e}^{-\sigma} \nonumber \\
        & + r_\mathrm{h}^{\sigma,-\sigma} \mathrm{e}^{\mathrm{i} k_{\mathrm{h}}^{-\sigma} z} \chi_\mathrm{h}^{-\sigma}
        + r_\mathrm{h}^{\sigma,\sigma} \mathrm{e}^{\mathrm{i} k_{\mathrm{h}}^{\sigma} z} \chi_\mathrm{h}^{\sigma}
        \label{eq:statesFM}
        \intertext{\normalsize in the ferromagnet and}
        \psi^{\sigma} (z>0) &= t_\mathrm{e}^{\sigma,\sigma} \mathrm{e}^{\mathrm{i} q_{\mathrm{e}} z} [u, 0, v, 0]^\top 
        + t_\mathrm{e}^{\sigma,-\sigma} \mathrm{e}^{\mathrm{i} q_{\mathrm{e}} z} [0, u, 0, v]^\top \nonumber \\
        & + t_\mathrm{h}^{\sigma,\sigma} \mathrm{e}^{-\mathrm{i} q_{\mathrm{h}} z} [v, 0, u, 0]^\top
        + t_\mathrm{h}^{\sigma,-\sigma} \mathrm{e}^{-\mathrm{i} q_{\mathrm{h}} z} [0, v, 0, u]^\top
        \label{eq:statesS}
    \end{align}
    }%
    in the superconductor. 
    The spinors for spin-$ \sigma $ electrons and holes in the ferromagnet are 
    {\small 
    \begin{align}
        \chi_\mathrm{e}^\sigma &= [\sigma \sqrt{1+\sigma \cos(\theta)} \mathrm{e}^{-\mathrm{i} \phi} , \sqrt{1-\sigma \cos(\theta)} , 0 , 0]^\top / \sqrt{2} 
        \intertext{\normalsize and}
        \chi_\mathrm{h}^\sigma &= [0, 0, -\sigma \sqrt{1-\sigma \cos(\theta)} \mathrm{e}^{-\mathrm{i} \phi} , \sqrt{1+\sigma \cos(\theta)}]^\top / \sqrt{2} ,
    \end{align}
    }
    respectively, while the Bardeen--Cooper--Schrieffer coherence factors in the superconductor at excitation energy~$ E > 0 $ fulfill 
    \begin{equation}
        u^2 = \frac{1}{2} \left( 1 + \frac{\sqrt{E^2 - \Delta_0^2}}{E} \right) = 1-v^2 
        \label{eq:bcs}
    \end{equation}
    Within Andreev approximation, we assume that $ E , \Delta_0 \ll \mu $ to approximate the electron(like) and hole(like) wave vectors as 
    \begin{align}
        k_{\mathrm{e}}^\sigma \approx k_{\mathrm{h}}^\sigma &\approx \sqrt{k_\mathrm{F}^2 (1+\sigma P) - \mathbf{k}_\mathbf{\parallel}^2} 
        \label{eq:WVferro}
        \intertext{in the ferromagnet and} 
        q_{\mathrm{e}} \approx q_{\mathrm{h}} &\approx \sqrt{k_\mathrm{F}^2 - \mathbf{k}_\mathbf{\parallel}^2}
        \label{eq:WVsuper}
    \end{align}
    in the superconductor; $ P = (\Delta_\mathrm{XC}/2)/\mu $ is a measure for the spin polarization of the ferromagnet and $ k_\mathrm{F} = \sqrt{2m\mu}/\hbar $ denotes the Fermi wave vector. 

    The energy-dependent scattering coefficients $ r_\mathrm{e}^{\sigma,(-)\sigma} (E) $, $ r_\mathrm{h}^{\sigma,(-)\sigma} (E) $, and $ t_\mathrm{e [h]}^{\sigma,(-)\sigma} (E) $---corresponding to (spin-flip) spin-conserving specular reflections, (spin-conserving) spin-flip Andreev reflections, and (spin-flip) spin-conserving electronlike [holelike] transmissions---are obtained applying the interfacial~(at $ z=0 $) boundary conditions
    \begin{equation}
        \psi^{\sigma} (z=0_-) = \psi^{\sigma} (z=0_+)
        \label{eq:boundary1}
    \end{equation}
    and 
    \begin{multline}
        \left\{ \left[ \frac{\hbar^2}{2m} \frac{\mathrm{d}}{\mathrm{d}z} + V_\mathrm{B} d_\mathrm{B} \right] \right\} \hat{\eta} \psi^{\sigma}(z) \big|_{z=0_-} \\
        + \left[ \begin{matrix} \hat{\boldsymbol{\Omega}} \cdot \hat{\boldsymbol{\sigma}} & \hat{0}_{2 \times 2} \\ \hat{0}_{2 \times 2} & -(\hat{\boldsymbol{\Omega}} \cdot \hat{\boldsymbol{\sigma}}) \end{matrix} \right] \psi^{\sigma}(z) \big|_{z=0_-} \\
        = \frac{\hbar^2}{2m} \frac{\mathrm{d}}{\mathrm{d} z} \hat{\eta} \psi^{\sigma}(z) \big|_{z=0_+} 
        \label{eq:boundary2}
    \end{multline}
    to the scattering states in~Eqs.~\eqref{eq:statesFM} and \eqref{eq:statesS}, and numerically solving the resulting system of equations~($ \hat{\eta} = \mathrm{diag}[\hat{\sigma}_0,-\hat{\sigma}_0] $). 

    \begin{widetext}

    Generalizing the Blonder--Tinkham--Klapwijk approach~\cite{Blonder1982}, the differential tunneling conductance at zero temperature is then evaluated from 
    \begin{equation}
        G_z = \frac{\mathrm{d} I_z}{\mathrm{d} V} = \frac{G_\mathrm{S}}{2\pi k_\mathrm{F}^2} \sum_{\sigma} \int \mathrm{d}^2 \mathbf{k_\parallel} \, \Bigg\{  1 - \mathrm{Re} \left[ \big| r_\mathrm{e}^{\sigma,\sigma} (eV) \big|^2 + \frac{k_{\mathrm{e}}^{-\sigma}}{k_{\mathrm{e}}^{\sigma}} \big| r_\mathrm{e}^{\sigma,-\sigma} (eV) \big|^2 \right] 
        + \mathrm{Re} \left[ \frac{k_{\mathrm{h}}^{-\sigma}}{k_{\mathrm{e}}^{\sigma}} \big| r_\mathrm{h}^{\sigma,-\sigma} (-eV) \big|^2 + \frac{k_{\mathrm{h}}^{\sigma}}{k_{\mathrm{e}}^{\sigma}} \big| r_\mathrm{h}^{\sigma,\sigma} (-eV) \big|^2 \right] \Bigg\} 
        \label{eq:btk}
    \end{equation}
    and the transverse tunneling-anomalous-Hall-effect conductances (computed in the ferromagnet close to the interface) from~\cite{Costa2019} 
    \begin{equation}
        G_{x (y)} = \frac{\mathrm{d} I_{x (y)}}{\mathrm{d} V} = - \frac{G_\mathrm{S}}{2\pi k_\mathrm{F}^2} \sum_{\sigma} \int \mathrm{d}^2 \mathbf{k_\parallel} \, \frac{k_{x (y)}}{k_{\mathrm{e}}^{\sigma}} \Bigg\{  \mathrm{Re} \left[ \big| r_\mathrm{e}^{\sigma,\sigma} (eV) \big|^2 + \big| r_\mathrm{e}^{\sigma,-\sigma} (eV) \big|^2 \right] 
        + \mathrm{Re} \left[ \big| r_\mathrm{h}^{\sigma,-\sigma} (-eV) \big|^2 + \big| r_\mathrm{h}^{\sigma,\sigma} (-eV) \big|^2 \right] \Bigg\} ;
        \label{eq:tahe}
    \end{equation}
    $ G_\mathrm{S} = A e^2 k_\mathrm{F}^2 / (2\pi h) $ denotes Sharvin's conductance of a perfectly transparent three-dimensional point contact~($ e $ is the positive elementary charge)~\footnote{For simplicity, the Hall-contact and interfacial cross-section areas are assumed to be equal and denoted by $ A $.}, 
    while taking the real parts~$ \mathrm{Re} (\ldots) $ ensures to only include propagating modes. 
    The transverse supercurrent responses~$ J_{x (y)} $ in the superconductor are computed from Green's functions analogously to Ref.~\cite{Costa2020}. 
    The technical details are summarized in Appendix~\ref{sec:AppA}. 

    \end{widetext}

    \section{Results  \label{sec:3}}

    To analyze the ramifications of radial Rashba SOC on superconducting transport, we numerically evaluate the tunneling and tunneling-anomalous-Hall-effect conductances by means of Eqs.~\eqref{eq:btk} and \eqref{eq:tahe} for realistic parameters. 
    The spin polarization of the ferromagnet is set to $ P = (\Delta_\mathrm{XC}/2) / \mu = 0.4 $, which corresponds to a weak ferromagnet~(for iron, $ P = 0.7 $), while the tunneling barrier is characterized by the dimensionless Blonder--Tinkham--Klapwijk $ Z $ parameter~\cite{Blonder1982} $ Z = 2m V_\mathrm{B} d_\mathrm{B} / (\hbar^2 k_\mathrm{F}) = 1 $, suggesting a high interfacial transparency of about $ \tau = [ 1+(Z/2)^2]^{-1} = 80 \, \% $; Rashba SOC $ \lambda_\mathrm{R} = 2m \alpha / \hbar^2 = 1 $ is the dominant SOC~(Dresselhaus SOC is $ \lambda_\mathrm{D} = 2m \beta / \hbar^2 = 0.1 $ in Figs.~\ref{fig:3} and \ref{fig:8}, tuned in Figs.~\ref{fig:4} and \ref{fig:7}, and absent otherwise) and we focus, for simplicity, on the zero-bias regime~($ eV = 0 $). 

    \subsection{Magnetoanisotropic Tunneling Transport}

\begin{figure}
        \centering
        \includegraphics[width=0.475\textwidth]{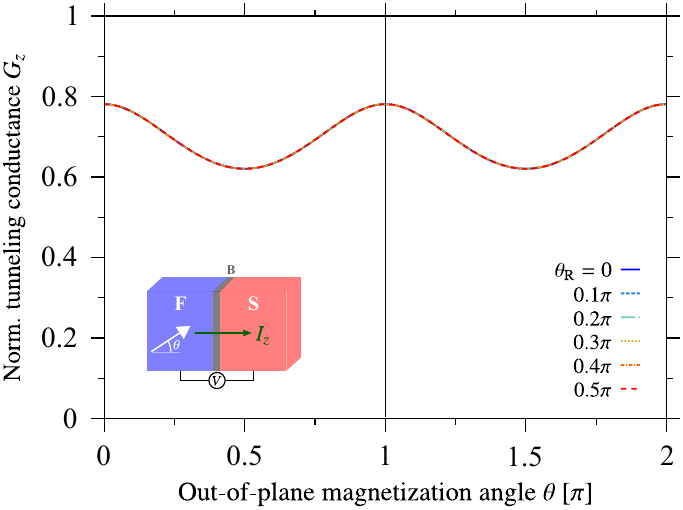}
        \caption{Calculated dependence of the tunneling conductance~$ G_z = \mathrm{d} I_z / \mathrm{d} V $---normalized to Sharvin's conductance $ G_\mathrm{S} = A e^2 k_\mathrm{F}^2 / (2\pi h) $---on the out-of-plane-plane magnetization angle~$ \theta $ for various indicated Rashba angles~$ \theta_\mathrm{R} \in [0;0.5\pi] $; the Rashba SOC parameter is $ \lambda_\mathrm{R} = 2m \alpha / \hbar^2 = 1 $ and Dresselhaus SOC is absent. }
        \label{fig:2}
    \end{figure}

    Tunneling magnetoanisotropies are among the most commonly used transport measures to probe interfacial SOC~\cite{Moser2007,Martinez2020}. 
    If only Rashba SOC is present, the in-plane magnetoanisotropy will disappear due to the in-plane invariance of the Rashba field, whereas the out-of-plane magnetoanisotropy will be finite and originate from a SOC-induced energy splitting of the ferromagnet's spin subbands~\cite{MatosAbiague2009}. 
    As this splitting is proportional to the strength $ \alpha $ of the Rashba SOC~\cite{Chantis2007} only, but independent of the Rashba angle~$ \theta_\mathrm{R} $, we predict that conventional and radial Rashba SOCs cannot be disentangled from out-of-plane magnetoanisotropy measurements. 
    Our model calculations presented in Fig.~\ref{fig:2} indeed confirm this expected $ \theta_\mathrm{R} $-invariance of the out-of-plane magnetoanisotropy.

    \begin{figure}
        \centering
        \includegraphics[width=0.475\textwidth]{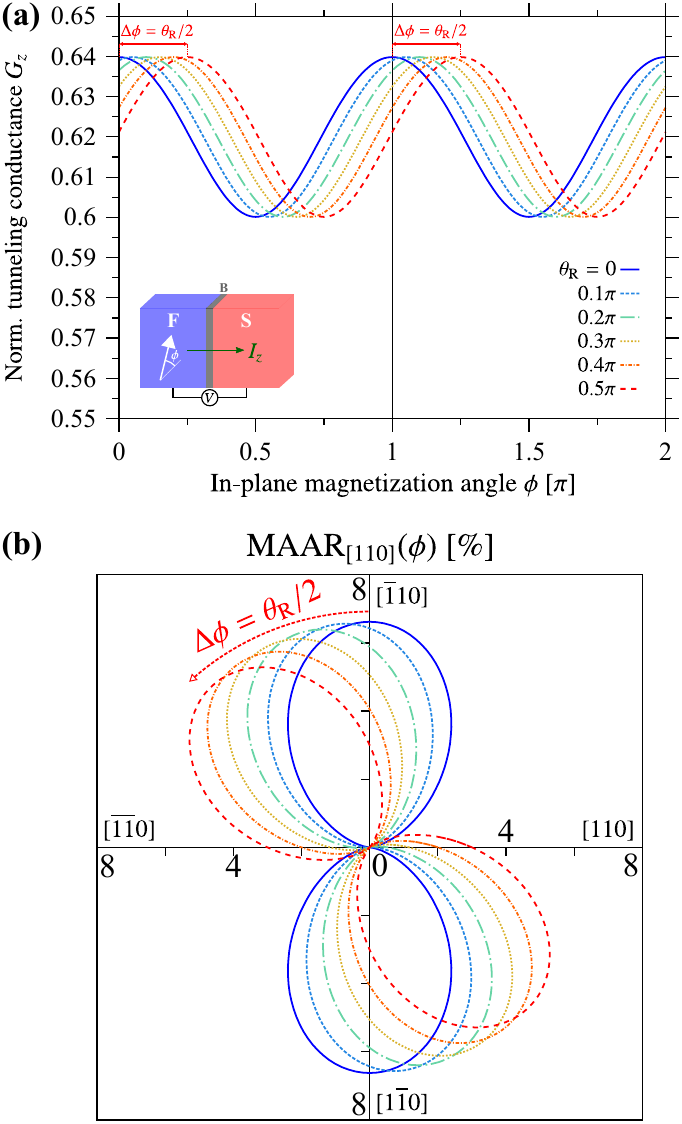}
        \caption{(a)~Calculated dependence of the tunneling conductance~$ G_z = \mathrm{d} I_z / \mathrm{d} V $---normalized to Sharvin's conductance $ G_\mathrm{S} = A e^2 k_\mathrm{F}^2 / (2\pi h) $---on the in-plane magnetization angle~$ \phi $ for various indicated Rashba angles~$ \theta_\mathrm{R} \in [0;0.5\pi] $; the Rashba~(Dresselhaus) SOC parameters are $ \lambda_\mathrm{R} = 2m \alpha / \hbar^2 = 1 $~($ \lambda_\mathrm{D} = 2m \beta / \hbar^2 = 0.1 \ll \lambda_\mathrm{R} $). (b)~Angular dependence of the in-plane magnetoanisotropic Andreev reflection (MAAR), evaluated according to~Eq.~(\ref{eq:maar}). }
        \label{fig:3}
    \end{figure}

    \begin{figure}
        \centering
        \includegraphics[width=0.475\textwidth]{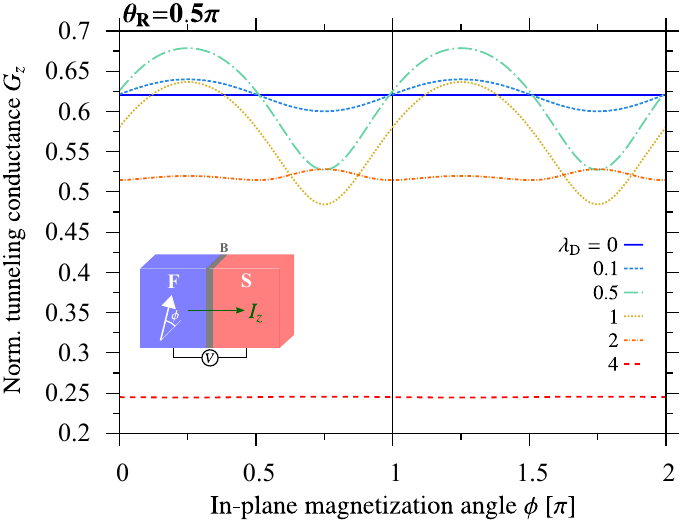}
        \caption{Calculated dependence of the tunneling conductance~$ G_z = \mathrm{d} I_z / \mathrm{d} V $---normalized to Sharvin's conductance $ G_\mathrm{S} = A e^2 k_\mathrm{F}^2 / (2\pi h) $---on the in-plane magnetization angle~$ \phi $ at purely radial Rashba SOC of strength $ \lambda_\mathrm{R} = 2m \alpha / \hbar^2 = 1 $ ($ \theta_\mathrm{R} = 0.5\pi $) interfering with various indicated Dresselhaus SOCs $ \lambda_\mathrm{D} = 2m \beta / \hbar^2 $. }
        \label{fig:4}
    \end{figure}

    To distinguish between conventional and radial Rashba SOCs from tunneling transport, we therefore need to focus on in-plane magnetoanisotropies~(indicating $ \theta = 0.5\pi $) in the simultaneous presence of Rashba and Dresselhaus SOCs; Dresselhaus SOC has not yet been conclusively identified in twisted van der Waals materials but a small contribution could emerge, e.g., from straining at multilayer interfaces~\cite{Bonell2010}. 
    For our calculations, we assume a hypothetic Dresselhaus SOC that is ten times weaker than the Rashba SOC, i.e., $ \lambda_\mathrm{D} = 0.1 \ll \lambda_\mathrm{R} $. 
    The in-plane orientation of the magnetization could be experimentally controlled either following the protocol given in~Refs.~\cite{Martinez2020,Tuero2024} or based on dysprosium magnets that remain magnetized along the pre-defined direction even after removing external magnetic fields~\cite{Betthausen2012}. 
    The calculated dependence of the tunneling conductance $ G_z $ on the in-plane magnetization angle~$ \phi $ in the ferromagnet is illustrated for different Rashba angles~$ \theta_\mathrm{R} \in [0;0.5\pi] $ in~Fig.~\ref{fig:3}(a). 
    As analyzed in~Ref.~\cite{Hoegl2015,*Hoegl2015a}, spin-flip Andreev reflections are the main source of transport magnetoanisotropies at subgap bias voltages. 
    The conductance contribution of these spin-flip Andreev reflections is maximal whenever the interfacial spin-orbit fields favor an in-plane spin orientation that is perpendicular to the preferred spin orientation (magnetization direction) in the ferromagnet---maximizing the probabilities for spin-flip scattering owing to the torque that acts on incoming electrons' spins---and minimal if both are (anti)parallel. 
    For conventional Rashba SOC ($ \theta_\mathrm{R} = 0 $), this results in maximal~(minimal) conductance for in-plane magnetization angles $ \phi = 0 \, \mathrm{mod} \, \pi $ ($ \phi = 0.5 \pi \, \mathrm{mod} \, \pi$), whereas radial Rashba SOC ($ \theta_\mathrm{R} = 0.5\pi $) requires $ \phi = 0.25 \pi \, \mathrm{mod} \, \pi $ ($ \phi = 0.75 \pi \, \mathrm{mod} \, \pi $); recall the orientation of the corresponding spin-orbit fields relative to the magnetization direction shown in Figs.~\ref{fig:1}(d) and \ref{fig:1}(f). 
    A general admixture of conventional and radial Rashba SOCs, quantified by the Rashba angle $ \theta_\mathrm{R} $, imprints thus an overall $ \Delta \phi = \theta_\mathrm{R}/2 $ shift on the angular dependence of the in-plane conductance magnetoanisotropy that is even more clearly evident when investigating the in-plane magnetoanisotropic Andreev reflection~\footnote{Note that we generalized the formula for the in-plane magnetoanisotropic Andreev reflection stated in Ref.~\cite{Hoegl2015,*Hoegl2015a}. To obtain the maximal amplitudes of the magnetoanisotropy, the magnetoanisotropic Andreev reflection needs to be computed with respect to $ G_z(\theta,\theta_\mathrm{R}/2) $, which corresponds to the maximal conductance at general Rashba angle~$ \theta_\mathrm{R} $ as shown in~Fig.~\ref{fig:3}(a). For conventional Rashba SOC, $ \theta_\mathrm{R} = 0 $ and the reference conductance is simply $ G_z(\theta,0) $ as used in~Ref.~\cite{Hoegl2015,*Hoegl2015a}.}, which is computed from 
    \begin{equation}
        \mathrm{MAAR}_{[110]} (\phi) = \frac{G_z(\theta,\theta_\mathrm{R}/2) - G_z(\theta,\phi)}{G_z(\theta,\phi)} \Bigg|_{\theta=0.5\pi} 
        \label{eq:maar}
    \end{equation} 
    and shown in Fig.~\ref{fig:3}(b).

    Figure~\ref{fig:4} illustrates the role of Dresselhaus SOC when interfering with purely radial Rashba SOC~($ \theta_\mathrm{R}=0.5\pi $) further. The Dresselhaus parameter is consecutively increased from $ \lambda_\mathrm{D} = 0 $ to $ 4 $, while the Rashba strength $ \lambda_\mathrm{R} = 1 $ is kept constant. As mentioned above, the tunneling conductance $ G_z $ is invariant under in-plane rotations of the magnetization if only Rashba SOC alone is present~($ \lambda_\mathrm{D} = 0 $, blue curve). Slowly increasing $ \lambda_\mathrm{D} $ to $ \lambda_\mathrm{D} \to \lambda_\mathrm{R} = 1 $, we recover the typically with radial Rashba SOC associated in-plane-magnetization dependence with maximal $ G_z $ at magnetization angles $ \phi = 0.25 \pi \, \mathrm{mod} \, \pi $, as analyzed in~Fig.~\ref{fig:3}. We note moreover that the magnetoanisotropic change of the conductance amplitudes---i.e., the magnetoanisotropic Andreev reflection---becomes maximal as the Dresselhaus parameter approaches the strength of the radial Rashba SOC~($ \lambda_\mathrm{D} \to \lambda_\mathrm{R} = 1 $). The physical reason can be deduced from the qualitative interference of the spin-orbit fields, as shown in~Fig.~\ref{fig:1}(f). As $ \lambda_\mathrm{D} \to \lambda_\mathrm{R} $, the Dresselhaus spin-orbit field fully compensates the oppositely aligned components of the radial Rashba field parallel to the diagonal $ [010] $-direction. As a consequence, if the magnetization in the ferromagnet is perpendicular to $ [010] $, i.e., $ \phi = 0.75\pi \, \mathrm{mod} \, \pi $, incoming electrons will not be exposed to any SOC perpendicular to their spin and the spin-flip Andreev-reflection contributions to the tunneling conductance will be strongly suppressed. The total conductance $ G_z $ therefore drops into a local minimum at $ \phi = 0.75\pi \, \mathrm{mod} \, \pi $~(see light-brown curve in Fig.~\ref{fig:4}), explaining the sizable relative difference between the maximal $ G_z $ at $ \phi = 0.25 \pi \, \mathrm{mod} \, \pi $ and its minima at $ \phi = 0.75 \pi \, \mathrm{mod} \, \pi $~(which results in magnetoanisotropic-Andreev-reflection amplitudes up to $ 25 \, \% $). The $ \Delta \phi = \theta_\mathrm{R} / 2 $ shift of $ G_z (\phi) $ that we proposed to identify radial Rashba SOC in experiments is still robust against these moderate Dresselhaus SOC strengths and only disappears if Dresselhaus becomes the dominant SOC mechanism, $ \lambda_\mathrm{D} \gtrsim 2 > \lambda_\mathrm{R} $, which is extremely unlikely to happen in twisted van der Waals barriers for which no signatures of Dresselhaus SOC have been found so far. The interplay between Dresselhaus and conventional Rashba SOCs has been analyzed in detail together with the in-plane magnetoanisotropic Andreev reflection in~Ref.~\cite{Hoegl2015,*Hoegl2015a}.

    \subsection{Tunneling Anomalous Hall Effect}

    How to distinguish radial from conventional Rashba SOC in the absence of Dresselhaus SOC? 
    We previously proposed that spin- and transverse-momentum-dependent skew scattering (filtering) of incoming spin-polarized electrons at SOC-inducing barriers additionally raises a tunneling-anomalous-Hall-effect response, which is sizable in magnitude in superconducting junctions owing to a constructive interference of skew specular and Andreev reflections~\cite{Costa2019}. 
    In contrast to the above-studied tunneling conductance, the tunneling anomalous Hall effect is of first order in SOC and reflects a nontrivial dependence on the in-plane magnetization angle $ \phi $ already if only Rashba spin-orbit fields alone are present~\cite{MatosAbiague2015}. The non-collinear spin quantization axes, arising from the interplay of magnetization and spin-orbit fields, cause the in-plane magnetoanisotropy of the tunneling anomalous Hall effect. 
    
    The following qualitative argument~\cite{MatosAbiague2015,Costa2019} elucidates the mechanism of this magnetoanisotropy. For a fixed $ \mathbf{k_\parallel} $-channel, the two preferred spatial directions in the junction are defined by the magnetization direction $ \hat{\mathbf{m}} $ and the spin-orbit field $ \hat{\boldsymbol{\Omega}} (\mathbf{k_\parallel}) $. 
    We can therefore expand the scattering-reflection-dependent part of the tunneling-anomalous-Hall-effect conductance formula, Eq.~\eqref{eq:tahe}, into the power series 
    \begin{equation}
        G_{x (y)} \approx \frac{G_\mathrm{S}}{2\pi k_\mathrm{F}^2} \sum_{\sigma} \sum_{n=1}^\infty \int \mathrm{d}^2 \mathbf{k_\parallel} \, \frac{k_{x (y)}}{k_{\mathrm{e}}^{\sigma}} \, c_n^\sigma \, \big[ \hat{\mathbf{m}} \cdot \hat{\boldsymbol{\Omega}} (\mathbf{k_\parallel}) \big]^n 
    \end{equation}
    with (in general complex) expansion coefficients $ c_n^\sigma $. 
    Scrutinizing the parity of the $ \mathbf{k_\parallel} $ integrand, we conclude that both $ G_{x (y)} $ are in first order approximated by the power-series contribution for $ n=1 $~(the contribution for $ n=0 $ vanishes due to parity), yielding 
    \begin{align}
        G_x &\propto - \alpha \sin (\phi - \theta_\mathrm{R})
        \intertext{and}
        G_y &\propto \phantom{-} \alpha \cos (\phi - \theta_\mathrm{R})
    \end{align}
    for the in-plane magnetization-angle dependence of the tunneling-anomalous-Hall-effect conductances. 
    Note that this result generalizes our findings in~Ref.~\cite{Costa2019}, in which only conventional Rashba SOC was present corresponding to $ \theta_\mathrm{R} = 0 $. 
    \emph{Probing the anisotropy of the tunneling-anomalous-Hall-effect conductances under in-plane magnetization rotations is thus predicted to enable the determination of $ \theta_\mathrm{R} $} even in the absence of in-plane spin-orbit anisotropies.  
    
    We wish to emphasize that similar physics arises also in normal-conducting junctions, for which first experiments have already successfully probed the tunneling anomalous Hall effect in nanogranular films~\cite{Rylkov2017}; the advantage of superconducting junctions is that additional skew Andreev reflections substantially enhance the tunneling anomalous Hall effect and produce a more robust signal, in agreement with our numerical simulations.

    \begin{figure}
        \centering
        \includegraphics[width=0.475\textwidth]{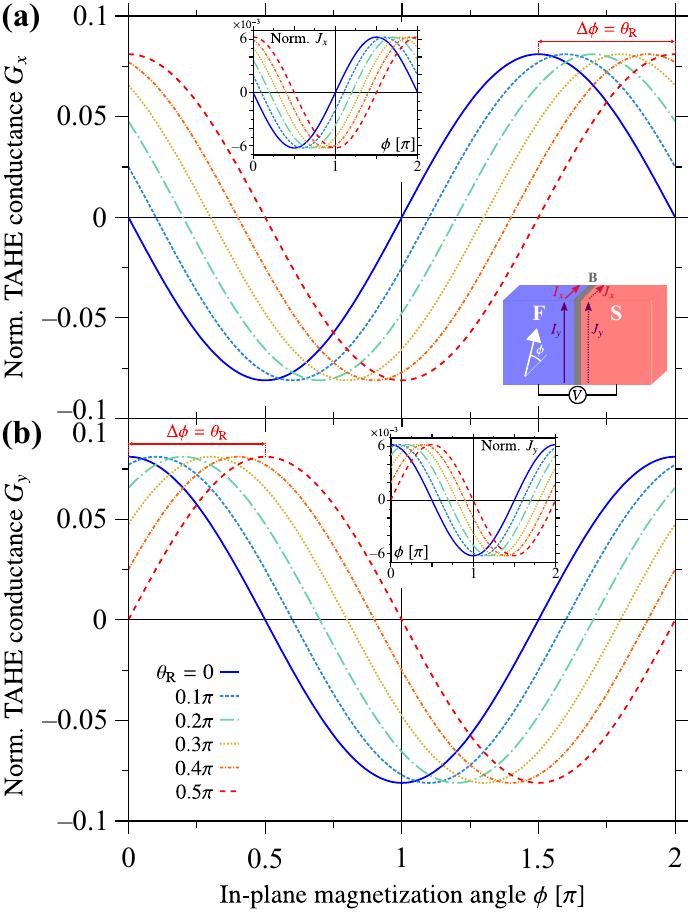}
        \caption{Calculated tunneling-anomalous-Hall-effect~(TAHE) conductances (a)~$ G_x = \mathrm{d} I_x / \mathrm{d} V $ and (b)~$ G_y = \mathrm{d} I_y / \mathrm{d} V $---normalized to Sharvin's conductance $ G_\mathrm{S} = A e^2 k_\mathrm{F}^2 / (2\pi h) $---as functions of the in-plane magnetization angle~$ \phi $ and for various indicated Rashba angles~$ \theta_\mathrm{R} \in [0;0.5\pi] $ at Rashba SOC $ \lambda_\mathrm{R} = 2m \alpha / \hbar^2 = 1 $; Dresselhaus SOC is absent. The insets show the corresponding Hall supercurrent responses $ J_x $ and $ J_y $, computed in the superconductor and given in multiples of $ \pi \Delta_0 G_\mathrm{S} / e $. }
        \label{fig:5}
    \end{figure}

\begin{figure}
        \centering
        \includegraphics[width=0.475\textwidth]{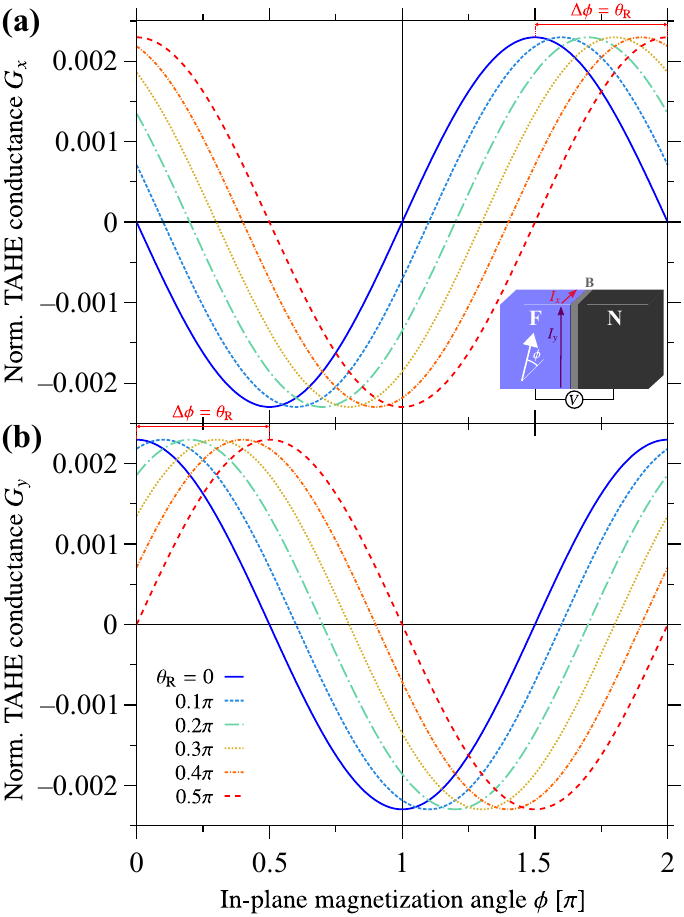}
        \caption{Calculated \emph{normal-state} (corresponding to the large-bias limit $ eV \gg \Delta_0 $) tunneling-anomalous-Hall-effect~(TAHE) conductances (a)~$ G_x = \mathrm{d} I_x / \mathrm{d} V $ and (b)~$ G_y = \mathrm{d} I_y / \mathrm{d} V $---normalized to Sharvin's conductance $ G_\mathrm{S} = A e^2 k_\mathrm{F}^2 / (2\pi h) $---as functions of the in-plane magnetization angle~$ \phi $ and for various indicated Rashba angles~$ \theta_\mathrm{R} \in [0;0.5\pi] $ at Rashba SOC $ \lambda_\mathrm{R} = 2m \alpha / \hbar^2 = 1 $; Dresselhaus SOC is absent. }
        \label{fig:6}
    \end{figure}

    Figure~\ref{fig:5} presents the tunneling-anomalous-Hall-effect conductances $ G_x $ and $ G_y $ as functions of the in-plane magnetization angle $ \phi $ and for different Rashba angles $ \theta_\mathrm{R} \in [0 ; 0.5\pi] $. 
    Both {Hall} conductances reflect the $ \phi $ dependencies that we deduced from parity arguments with the $ \Delta \phi = \theta_\mathrm{R} $ shifts at general $ \theta_\mathrm{R} $. 
    The maximal amplitudes of {$ G_x $ and $ G_y $} are independent of $ \theta_\mathrm{R} $, as we assumed the same SOC amplitude $ \alpha $ for conventional and radial Rashba-SOC contributions. 
    Since there is no Dresselhaus SOC to interfere with, the \emph{maximal} {Hall} conductances are not subject to any crystallographic anisotropy, and their magnitudes are equal along the $ \hat{x} $- and $ \hat{y} $-directions. 
    Moreover, we note the sizable amplitudes of the Hall conductances~(reaching nearly $ 10 \, \% $ of the respective tunneling conductance) due to the aforementioned skew Andreev reflections in the superconducting state; in the normal state, our calculations yield similar physics~(i.e., the same $ \Delta \phi = \theta_\mathrm{R} $ shifts from which the Rashba angle can be experimentally extracted), but the absolute amplitudes of $ G_x $ and $ G_y $ are up to two orders of magnitude smaller (and less than $ 1 \, \% $ of the respective tunneling conductance) as shown in~Fig.~\ref{fig:6}, respectively. 
    
    As we demonstrated in~Ref.~\cite{Costa2019}, the skew Andreev-reflection process cycles Cooper pairs to the superconducting side of the junction, which are undergoing the spin- and transverse-momentum-dependent filtering as well, and trigger Hall supercurrent responses denoted by $ J_x $ and $ J_y $. 
    The $ \phi $ dependencies of $ J_x $ and $ J_y $, calculated from our model as outlined in Appendix~\ref{sec:AppA} and presented in the insets of Fig.~\ref{fig:5}, show the same $ \Delta \phi = \theta_\mathrm{R} $ shifts as the tunneling-anomalous-Hall-effect conductances, which convinces us that both phenomena are indeed closely connected and result from the same skew-scattering mechanism.

\begin{figure}
        \centering
        \includegraphics[width=0.475\textwidth]{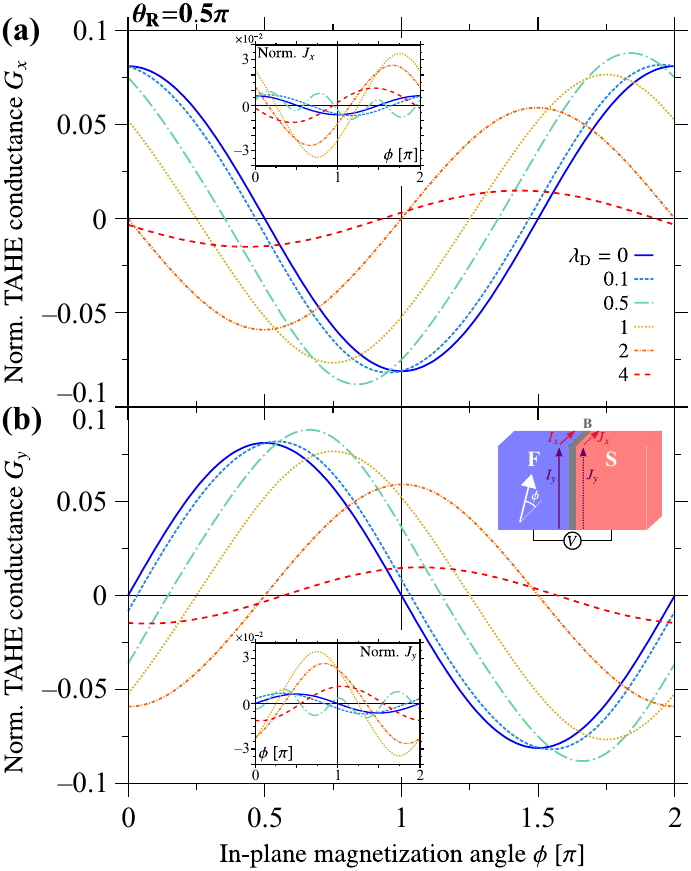}
        \caption{Calculated tunneling-anomalous-Hall-effect~(TAHE) conductances (a)~$ G_x = \mathrm{d} I_x / \mathrm{d} V $ and (b)~$ G_y = \mathrm{d} I_y / \mathrm{d} V $---normalized to Sharvin's conductance $ G_\mathrm{S} = A e^2 k_\mathrm{F}^2 / (2\pi h) $---as functions of the in-plane magnetization angle~$ \phi $ at purely radial Rashba SOC of strength $ \lambda_\mathrm{R} = 2m \alpha / \hbar^2 = 1 $~($ \theta_\mathrm{R} = 0.5\pi $) interfering with various indicated Dresselhaus SOCs~$ \lambda_\mathrm{D} = 2m \beta / \hbar^2 $. The insets show the corresponding Hall supercurrent responses $ J_x $ and $ J_y $, computed in the superconductor and given in multiples of $ \pi \Delta_0 G_\mathrm{S} / e $. }
        \label{fig:7}
    \end{figure}

    Another advantage of the detection of radial Rashba SOC through tunneling-anomalous-Hall-effect measurements is the possibility to still disentangle its contribution from Dresselhaus SOC in situations in which also the latter is relevant. 
    While radial Rashba and Dresselhaus spin-orbit fields have been predicted to manifest in the same magnetotransport and supercurrent-diode-effect characteristics in lateral (graphene-based) heterostructures~\cite{Kang2024}---owing to their coupling to the wave vector parallel to the tunneling current, which is similar for radial Rashba and Dresselhaus components---their coupling to wave vectors perpendicular to the tunneling direction gives rise to well-distinct ramifications in the vertical junctions considered in this paper. 
    Generalizing the parity arguments from above allowing also for nonzero Dresselhaus SOC~(parametrized by $ \beta $), we obtain 
    \begin{align}
        G_x &\propto -\alpha \sin(\phi - \theta_\mathrm{R}) - \beta \sin(\phi) 
        \intertext{and}
        G_y &\propto \alpha \cos(\phi - \theta_\mathrm{R}) - \beta \cos(\phi) . 
    \end{align}
    The $ \theta_\mathrm{R} $ shift of $ G_x $~($ G_y $) is therefore fully induced by the radial Rashba component alone, while Dresselhaus SOC produces mainly an ``offset'' with the usual sinusoidal~(co-sinusoidal) magnetization-angle dependence. 
    As shown in Fig.~\ref{fig:7} for purely radial Rashba SOC~($ \theta_\mathrm{R} = 0.5\pi $), the $ \theta_\mathrm{R} $ shifts---and thereby the presence of radial Rashba SOC---are clearly resolvable from the tunneling anomalous Hall effect even for Dresselhaus parameters approaching the strength of the Rashba SOC~($ \beta \to \alpha $); the amplitudes of the shifts are, however, suppressed with stronger Dresselhaus SOC as the nonshifted $ \sin $~($ \cos $) terms start to dominate producing higher-harmonic terms in the total Hall conductances. 
    Only if $ \beta \gg \alpha $---which is extremely unlikely for the considered systems as mentioned before---the $ \theta_\mathrm{R} $ shifts (completely) disappear and the Hall conductances are not conclusive to identify radial Rashba SOC. 
    The \emph{maximal} amplitudes of the tunneling-anomalous-Hall-effect conductances are remarkably damped at strong total SOC, as the SOC terms in the deltalike model additionally enhance interfacial scattering~(reduce the transparency of the interface). 
    The behavior of the Hall supercurrent responses~$ J_x $ and $ J_y $ in the simultaneous presence of radial Rashba and Dresselhaus SOC~(see the insets of Fig.~\ref{fig:7}) is qualitatively similar to the Hall conductances. 
    However, we note that the higher-harmonic terms stemming from the interference of the shifted radial Rashba and nonshifted Dresselhaus contributions are more pronounced in the Hall supercurrents and that the amplitudes of $ J_{x (y)} $ initially remarkably increase with stronger Dresselhaus SOC~(before the final suppression due to the aforementioned additional scattering at the interface). 
    The latter observation might be a possible indication of rather sizable triplet components in the Hall supercurrents. 

    \section{More realistic junctions: Chemical-potential and mass mismatches  \label{sec:4}}
    
    For simplicity, we have so far assumed equal chemical potentials and effective masses---and thereby also equal Fermi wave vectors---in the ferromagnetic and superconducting junction regions, respectively. 
    As this is typically not the case in real junctions consisting of different materials as electrodes, we need to address the impact of chemical-potential and mass mismatches on the results discussed above.

    To account for different chemical potentials and masses, we modify the electron Hamiltonian $ \hat{\mathcal{H}}_\mathrm{e} $ in the Bogoliubov--de Gennes Hamiltonian, Eq.~\eqref{eq:BdG}, according to 
    \begin{align}
        \hat{\mathcal{H}}_\mathrm{e} &= \left( -\frac{\hbar^2}{2} \boldsymbol{\nabla} \frac{1}{m(z)} \boldsymbol{\nabla} - \mu(z) \right) \hat{\sigma}_0 \nonumber \\
        &\hspace{50 pt} - \frac{\Delta_\mathrm{XC}}{2} \, \Theta(-z) \, (\hat{\mathbf{m}} \cdot \hat{\boldsymbol{\sigma}} ) + \hat{\mathcal{H}}_\mathrm{B} ,
    \end{align}
    where 
    \begin{align}
        m(z) &= m_\mathrm{F} \, \Theta(-z) + m_\mathrm{S} \, \Theta(z)
        \intertext{and}
        \mu(z) &= \mu_\mathrm{F} \, \Theta(-z) + \mu_\mathrm{S} \, \Theta(z) ;
    \end{align}
    $ m_\mathrm{F (S)} $ indicates the effective electron mass and $ \mu_\mathrm{F (S)} $ is the chemical potential in the ferromagnet~(superconductor). 
    The solutions of the Bogoliubov--de Gennes equation $ \hat{\mathcal{H}}_\mathrm{BdG} \Psi^\sigma (\mathbf{r}) = E \Psi^\sigma (\mathbf{r}) $ are still given by Eqs.~\eqref{eq:statesFM} and \eqref{eq:statesS} taking the different masses and chemical potentials in the wave vectors into account such that 
    \begin{align}
        k_\mathrm{e}^\sigma &\approx k_\mathrm{h}^\sigma \approx \sqrt{k_\mathrm{F}^2 (1+\sigma P) - \mathbf{k}_\mathbf{\parallel}^2} 
        \intertext{and}
        q_\mathrm{e} &\approx q_\mathrm{h} \approx \sqrt{q_\mathrm{F}^2 - \mathbf{k}_\mathbf{\parallel}^2} ;
    \end{align}
    $ k_\mathrm{F} = \sqrt{2m_\mathrm{F} \mu_\mathrm{F}}/\hbar $~($ q_\mathrm{F} = \sqrt{2m_\mathrm{S} \mu_\mathrm{S}}/\hbar $) denotes the Fermi wave vector in the ferromagnet~(superconductor) within Andreev approximation [$ E, \Delta_0 \ll \mu_\mathrm{F (S)} $], while $ P = (\Delta_\mathrm{XC}/2)/\mu_\mathrm{F} $ measures the spin polarization of the ferromagnet. 
    The barrier and Rashba (Dresselhaus) SOC parameters in the presence of mass and Fermi-wave-vector mismatches read as $ Z = 2 \sqrt{m_\mathrm{F} m_\mathrm{S}} V_\mathrm{B} d_\mathrm{B} / (\hbar^2 \sqrt{k_\mathrm{F} q_\mathrm{F}}) $ and $ \lambda_\mathrm{R} = 2 \sqrt{m_\mathrm{F} m_\mathrm{S}} \alpha / \hbar^2 $ ($ \lambda_\mathrm{D} = 2 \sqrt{m_\mathrm{F} m_\mathrm{S}} \beta / \hbar^2 $)---which transform into the parameters used above substituting $ m_\mathrm{F} = m_\mathrm{S} \equiv m $ and $ \mu_\mathrm{F} = \mu_\mathrm{S} \equiv \mu $. 
    The mass~(Fermi-wave-vector to additionally cover different chemical potentials) mismatch is quantified by the dimensionless parameter $ F_\mathrm{M} = m_\mathrm{S} / m_\mathrm{F} $~($ F_\mathrm{K} = q_\mathrm{F} / k_\mathrm{F} $); note that a similar approach has already been applied to treat different Fermi levels in superconducting junctions earlier, e.g., in~Refs.~\cite{Zutic1999,Zutic2000,*Zutic2000alt}.

    \begin{figure}
        \centering
        \includegraphics[width=0.475\textwidth]{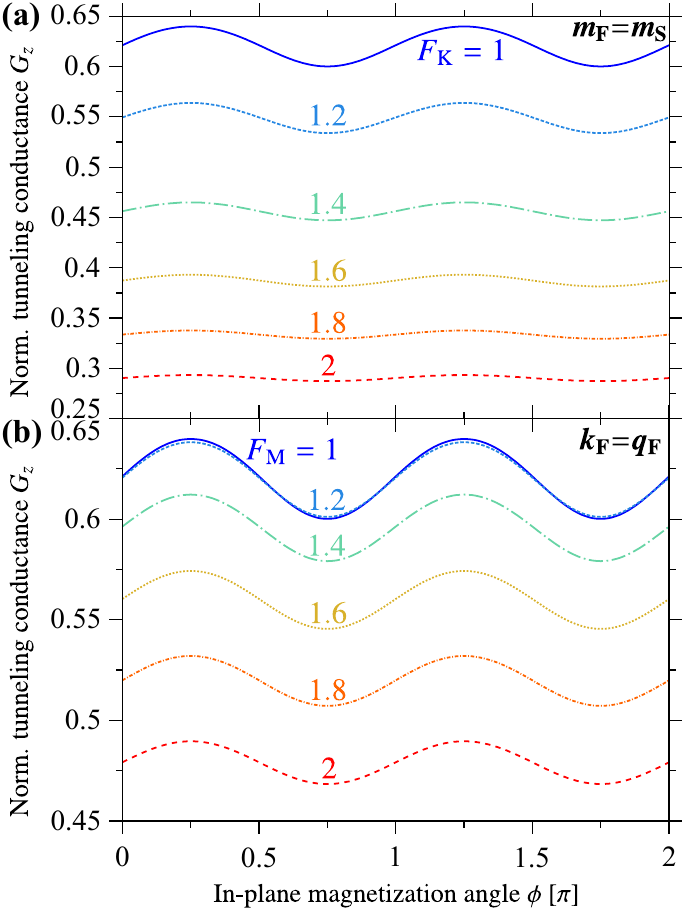}
        \caption{(a)~Calculated dependence of the tunneling conductance~$ G_z = \mathrm{d} I_z / \mathrm{d} V $---normalized to Sharvin's conductance $ G_\mathrm{S} = A e^2 k_\mathrm{F}^2 / (2\pi h) $---on the in-plane magnetization angle~$ \phi $ at purely radial Rashba SOC of strength $ \lambda_\mathrm{R} = 2 \sqrt{m_\mathrm{F} m_\mathrm{S}} \alpha / \hbar^2 = 1 $ ($ \theta_\mathrm{R} = 0.5\pi $) interfering with weak Dresselhaus SOC $ \lambda_\mathrm{D} = 2 \sqrt{m_\mathrm{F} m_\mathrm{S}} \beta / \hbar^2 = 0.1 $ for various indicated Fermi-wave-vector mismatches $ F_\mathrm{K} = q_\mathrm{F} / k_\mathrm{F} $ (assuming equal masses, $ F_\mathrm{M} = m_\mathrm{S} / m_\mathrm{F} = 1 $). (b)~Same calculation as in~(a) for various indicated mass mismatches $ F_\mathrm{M} = m_\mathrm{S} / m_\mathrm{F} $ at equal Fermi wave vectors ($ F_\mathrm{K} = q_\mathrm{F} / k_\mathrm{F} = 1 $). }
        \label{fig:8}
    \end{figure}

\begin{figure}
        \centering
        \includegraphics[width=0.475\textwidth]{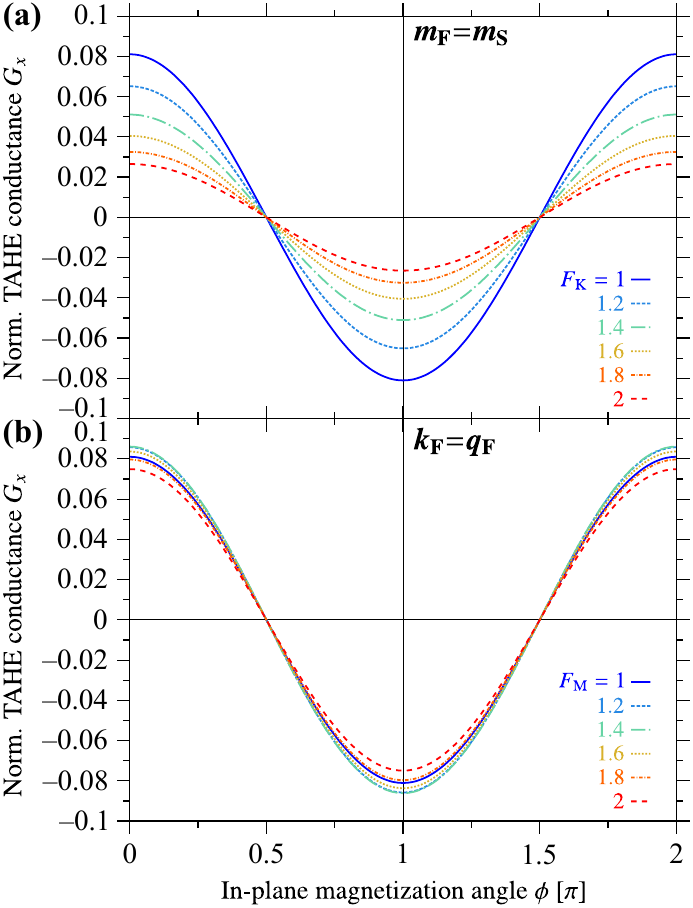}
        \caption{(a)~Calculated tunneling-anomalous-Hall-effect~(TAHE) conductance $ G_x = \mathrm{d} I_x / \mathrm{d} V $---normalized to Sharvin's conductance $ G_\mathrm{S} = A e^2 k_\mathrm{F}^2 / (2\pi h) $---as a function of the in-plane magnetization angle~$ \phi $ at purely radial Rashba SOC of strength $ \lambda_\mathrm{R} = 2 \sqrt{m_\mathrm{F} m_\mathrm{S}} \alpha / \hbar^2 = 1 $ ($ \theta_\mathrm{R} = 0.5\pi $) for various indicated Fermi-wave-vector mismatches $ F_\mathrm{K} = q_\mathrm{F} / k_\mathrm{F} $ (assuming equal masses, $ F_\mathrm{M} = m_\mathrm{S} / m_\mathrm{F} = 1 $); Dresselhaus SOC is absent. (b)~Same calculation as in~(a) for various indicated mass mismatches $ F_\mathrm{M} = m_\mathrm{S} / m_\mathrm{F} $ at equal Fermi wave vectors ($ F_\mathrm{K} = q_\mathrm{F} / k_\mathrm{F} = 1 $). } 
        \label{fig:9}
    \end{figure}

    To verify the robustness of our predictions in~Sec.~\ref{sec:3}, we analyze the tunneling conductance $ G_z $ and the $ x $-component of the Hall conductance, $ G_x $, at different Fermi-wave-vector and mass mismatch ratios in Figs.~\ref{fig:8} and \ref{fig:9}, respectively. 
    As our paper focuses on the $ \theta_\mathrm{R} $-dependent shifts to experimentally probe radial Rashba SOC, we consider fully radial Rashba fields~($ \theta_\mathrm{R} = 0.5\pi $) and keep all other system parameters (barrier, Rashba, and Dresselhaus strengths) the same as in Sec.~\ref{sec:3}. 
    Most importantly, both the $ \Delta \phi = \theta_\mathrm{R}/2 $- and $ \Delta \phi = \theta_\mathrm{R} $ shifts of the tunneling and Hall conductances are not affected by Fermi-wave-vector or mass mismatches, and provide therefore indeed a robust experimental way to detect radial Rashba SOC also in real junctions. 
    Stronger mismatches additionally reduce the interfacial transparency of the junction and therefore suppress the amplitudes of the tunneling conductance $ G_z $~(magnetoanisotropic Andreev reflection)---Fermi-wave-vector mismatch even faster than mass mismatch. 
    Regarding the Hall conductance, larger Fermi-wave-vector mismatch again remarkably reduces the amplitudes of $ G_x $, whereas mass mismatch has only a minor quantitative impact and can even initially slightly increase $ G_x $ most likely due to the skew-scattering mechanism proposed in~Ref.~\cite{Costa2019} that depends non-trivially on the barrier parameters.

    These calculations confirm that our predictions to extract radial Rashba SOC from Rashba-angle shifts in the magnetoanisotropy of the tunneling or Hall conductances apply to more realistic junctions as well. 
    In fact, the authors of~Ref.~\cite{Kang2024} argue that introducing a nonzero $ \theta_\mathrm{R} $ in the Rashba Hamiltonian corresponds to an in-plane rotation of the spin-orbit fields that must leave transport in the same plane---like the Hall currents---invariant under a counter-rotated magnetic-exchange field. 
    These arguments emphasize that the $ \theta_\mathrm{R} $ shifts of the Hall conductances are protected by the symmetry of the spin-orbit fields and thus robust not only against chemical-potential or mass mismatches but also against most other perturbative effects in real junctions~(such as charging, interface impurities, or small straining) as long as the underlying symmetry of the Rashba fields is not destroyed.

    \section{Conclusions  \label{sec:5}}

    In summary, we applied well-established theoretical frameworks to investigate the magnetoanisotropies of the tunneling and tunneling-anomalous-Hall-effect conductance spectra of superconducting magnetic junctions in the simultaneous presence of conventional and unconventional radial Rashba SOCs at the interface formed by a thin tunneling barrier~(realized, e.g., through a twisted van der Waals bilayer). 
    We demonstrated that a finite Rashba angle $ \theta_\mathrm{R} $ imprints a robust $ \Delta \phi = \theta_\mathrm{R} $ shift on the in-plane magnetization-angle dependence of the Hall conductances in the ferromagnetic electrode, as well as on the corresponding Hall supercurrent responses in the superconductor, from which $ \theta_\mathrm{R} $ could be quantified in Hall-transport experiments. 
    Both the out-of-plane and in-plane tunneling conductances, however, have been shown to be invariant of $ \theta_\mathrm{R} $ if only Rashba SOC alone is present~(as it is so far believed to be the case in twisted van der Waals bilayers) and only provide a distinction between conventional and radial Rashba SOCs when interfering with another, functionally distinct, spin-orbit field of, e.g., the Dresselhaus type. 
    Studying tunneling-conductance magnetoanisotropies is thus nevertheless still an important experimental step to explicitly exclude possible Dresselhaus SOC in junctions with van der Waals barriers although the presence of weak Dresselhaus SOC would only have a minor impact on the $ \theta_\mathrm{R} $ shifts of the Hall conductances.

    \appendix
    \section{Calculation of the Hall supercurrent responses  \label{sec:AppA}}

    To calculate the Hall supercurrent responses on the superconducting side, we adapt the Furusaki--Tsukada~\cite{Furusaki1991} Green's-function approach similarly to Ref.~\cite{Costa2019}. 
    For simplicity, we assume that the superconducting region spans the $ z < 0 $-half space, while the ferromagnet is located at $ z > 0 $. 
    Analogously to our formulation in the main text, the Bogoljubov--de Gennes Hamiltonian can then be written as 
    \begin{equation}
        \hat{\mathcal{H}}_\mathrm{BdG} = \left[ \begin{matrix} \hat{\mathcal{H}}_\mathrm{e} & \Delta_0 \, \Theta(-z) \, \hat{\sigma}_0 \\ \Delta_0 \, \Theta(-z) \, \hat{\sigma}_0 & -\hat{\sigma}_y \, \hat{\mathcal{H}}_\mathrm{e}^* \, \hat{\sigma}_y \end{matrix} \right] 
    \end{equation}
    with the single-electron Hamiltonian 
    \begin{equation}
        \hat{\mathcal{H}}_\mathrm{e} = \left( -\frac{\hbar^2}{2m} \boldsymbol{\nabla}^2 - \mu \right) \hat{\sigma}_0 - \frac{\Delta_\mathrm{XC}}{2} \, \Theta(z) \, \left( \hat{\mathbf{m}} \cdot \hat{\boldsymbol{\sigma}} \right) + \hat{\mathcal{H}}_\mathrm{B} ;
    \end{equation}
    the interface is described by 
    \begin{multline}
        \hat{\mathcal{H}}_\mathrm{B} = \bigg\{ V_\mathrm{B} d_\mathrm{B} \hat{\sigma}_0 \\
        + \alpha \, \big[ \cos(\theta_\mathrm{R}) \left( k_y \hat{\sigma}_x - k_x \hat{\sigma}_y \right) \\
        + \sin(\theta_\mathrm{R}) \left( k_x \hat{\sigma}_x + k_y \hat{\sigma}_y \right) \big] \\
        - \beta \left( k_y \hat{\sigma}_x + k_x \hat{\sigma}_y \right) \bigg\} \, \delta(z) , 
    \end{multline}
    accounting for the tunneling barrier with height~(width) $ V_\mathrm{B} $~($ d_\mathrm{B} $), conventional and radial Rashba SOCs (strength $ \alpha $ and Rashba angle $ \theta_\mathrm{R} $), as well as Dresselhaus SOC (strength $ \beta $).

    The general solution of the Bogoliubov--de Gennes equation $ \hat{\mathcal{H}}_\mathrm{BdG} \Psi^{(1)} (\mathbf{r}) = E \Psi^{(1)}(\mathbf{r}) $ for an incoming (1)~spin-up electronlike quasiparticle from the superconductor is found to read as 
    \begin{equation}
        \Psi^{(1)} (\mathbf{r}) = \psi^{(1)}(z) \, \mathrm{e}^{\mathrm{i} \, (k_x x + k_y y)} , 
    \end{equation}
    where 
    \begin{multline}
        \psi^{(1)}(z<0) = \mathrm{e}^{\mathrm{i} q_{\mathrm{e}} z} \left[ \begin{matrix} u \\ 0 \\ v \\ \phantom{\,} 0 \phantom{\,} \end{matrix} \right] 
        + \mathcal{A}^{(1)} \, \mathrm{e}^{- \mathrm{i} q_{\mathrm{e}} z} \left[ \begin{matrix} u \\ 0 \\ v \\ \phantom{\,} 0 \phantom{\,} \end{matrix} \right] 
        + \mathcal{B}^{(1)} \, \mathrm{e}^{- \mathrm{i} q_{\mathrm{e}} z} \left[ \begin{matrix} 0 \\ u \\ \phantom{\,} 0 \phantom{\,} \\ v \end{matrix} \right] \\
        + \mathcal{C}^{(1)} \, \mathrm{e}^{ \mathrm{i} q_{\mathrm{h}} z} \left[ \begin{matrix} v \\ 0 \\ u \\ \phantom{\,} 0 \phantom{\,} \end{matrix} \right] 
        + \mathcal{D}^{(1)} \, \mathrm{e}^{ \mathrm{i} q_{\mathrm{h}} z} \left[ \begin{matrix} 0 \\ v \\ \phantom{\,} 0 \phantom{\,} \\ u \end{matrix} \right] 
    \end{multline}
    and
    \begin{align}
        \psi^{(1)}(z>0) &= \mathcal{E}^{(1)} \, \mathrm{e}^{\mathrm{i} k_{\mathrm{e}}^{\sigma=1} z} \, \frac{1}{\sqrt{2}} \left[ \begin{matrix} \sqrt{1+\cos(\theta)} \, \mathrm{e}^{-\mathrm{i} \phi} \\ \sqrt{1-\cos(\theta)} \\ 0 \\ 0 \end{matrix} \right] \nonumber \\
        &\hspace{5 pt} + \mathcal{F}^{(1)} \, \mathrm{e}^{\mathrm{i} k_{\mathrm{e}}^{\sigma=-1} z} \, \frac{1}{\sqrt{2}} \left[ \begin{matrix} -\sqrt{1-\cos(\theta)} \, \mathrm{e}^{-\mathrm{i} \phi} \\ \sqrt{1+\cos(\theta)} \\ 0 \\ 0 \end{matrix} \right] \nonumber \\
        &\hspace{5 pt} + \mathcal{G}^{(1)} \, \mathrm{e}^{-\mathrm{i} k_{\mathrm{h}}^{\sigma=-1} z} \, \frac{1}{\sqrt{2}} \left[ \begin{matrix} 0 \\ 0 \\ \sqrt{1+\cos(\theta)} \, \mathrm{e}^{-\mathrm{i} \phi} \\ \sqrt{1-\cos(\theta)} \end{matrix} \right] \nonumber \\
        &\hspace{15 pt}+ \mathcal{H}^{(1)} \, \mathrm{e}^{-\mathrm{i} k_{\mathrm{h}}^{\sigma=1} z} \, \frac{1}{\sqrt{2}} \left[ \begin{matrix} 0 \\ 0 \\ -\sqrt{1-\cos(\theta)} \, \mathrm{e}^{-\mathrm{i} \phi} \\ \sqrt{1+\cos(\theta)} \end{matrix} \right] ;
    \end{align}
    the Bardeen--Cooper--Schrieffer coherence factors $ u $ and $ v $ are given by Eq.~\eqref{eq:bcs}, and the spin-$ \sigma $ wave vectors within Andreev approximation~($ E , \Delta_0 \ll \mu $) by Eqs.~\eqref{eq:WVferro} and \eqref{eq:WVsuper}, respectively. 

    The scattering states $ \Psi^{(2)} (\mathbf{r}) $, $ \Psi^{(3)} (\mathbf{r}) $, and $ \Psi^{(4)} (\mathbf{r}) $ for incident (2)~spin-down electronlike, (3)~spin-up holelike, and (4)~spin-down holelike quasiparticles are obtained in an analogous manner. 
    The calligraphically written scattering coefficients are numerically determined applying interfacial~($ z=0 $) boundary conditions as stated in Eqs.~\eqref{eq:boundary1} and \eqref{eq:boundary2}, and solving the resulting linear systems of equations. Particularly relevant are the spin-conserving Andreev-reflection coefficients $ \mathcal{C}^{(1)} $, $ \mathcal{D}^{(2)} $, $ \mathcal{A}^{(3)} $, and $ \mathcal{B}^{(4)} $, which provide the input to calculate the Hall supercurrent responses close to the interface from~\cite{Costa2019} 
    \begin{multline}
        J_{x (y)} = \frac{\pi \Delta_0 G_\mathrm{S}}{e} \frac{k_\mathrm{B} T}{2 \pi k_\mathrm{F}^2} \int \mathrm{d}^2 \mathbf{k_\parallel} \, \sum_{\omega_n} \frac{k_{x (y)}}{\sqrt{k_\mathrm{F}^2 - \mathbf{k}_\mathbf{\parallel}^2}} \\
        \times \left[ \frac{\mathcal{C}^{(1)} (\mathrm{i} \omega_n) + \mathcal{D}^{(2)} (\mathrm{i} \omega_n) + \mathcal{A}^{(3)} (\mathrm{i} \omega_n) + \mathcal{B}^{(4)} (\mathrm{i} \omega_n)}{\sqrt{\omega_n^2 + \Delta_0^2}} \right] ;
        \label{eq:HallSupercurrent}
    \end{multline}
    $ k_\mathrm{B} T $ is the thermal energy at temperature $ T $ (we finally consider $ T / T_\mathrm{c} = 0.1 $, where $ T_\mathrm{c} $ is the critical temperature of the superconductor) and $ \omega_n = (2n + 1) \pi k_\mathrm{B} T $, with integer $ n $, are the fermionic Matsubara frequencies.

    \begin{acknowledgments}
        A.C. and J.F. gratefully acknowledge funding by Deutsche Forschungsgemeinschaft (DFG, German Research Foundation) -- Project-IDs 454646522; 314695032. 
    \end{acknowledgments}

    \bibliography{paper}

\end{document}